\documentclass[11pt]{article} 
\usepackage[margin=1in]{geometry}
\usepackage{empheq} 
\usepackage{fontenc}
\usepackage{inputenc}
\usepackage{amssymb}
\usepackage{amsthm}
\usepackage{amsmath}
\usepackage{algorithmicx, algpseudocode}
\usepackage[linesnumbered,ruled,vlined]{algorithm2e}
\usepackage{graphicx}
\usepackage{subcaption }

\providecommand{\keywords}[1]
{
  \small	
  \textbf{\textit{Keywords---}} #1
}

\theoremstyle{definition}

\theoremstyle{remark}

\title{Cyberattack on the Microgrids Through Price Modification}
\author{ Subhankar Mishra \\
smishra@niser.ac.in
\\ 
School of Computer Sciences, NISER, \\Bhubaneswar India. \\ 
Homi Bhabha National Institute, \\ Anushaktinagar, Mumbai - 400094, India}
\date{}
\begin{document}
\maketitle

\abstract{
    Recent massive failures in the power grid acted as a wake up call for all utilities and consumers. This leads to aggressive pursue a more intelligent grid which addresses the concerns of reliability, efficiency, security, quality and sustainability for the energy consumers and producers alike. One of the many features of the smart grid is a discrete energy system consisting of distributed energy sources capable of operating independently from the main grid known as the microgrid. The main focus of the microgrid  is to ensure a reliable and affordable energy security. However, it also can be vulnerable to cyber attack and we study the effect of price modification of electricity attack on the microgrid, given that they are able to operate independently from the main grid. This attack consists of two stages, 1) Separate the microgrids from the main grid (islanding) and 2) Failing the nodes inside the microgrid. Empirical results on IEEE Bus data help us evaluate our approach under various settings of grid parameters.
}
\\ \bigskip

\section{Introduction}

Failures in the power grid in the recent times with examples 2003 blackout in the Northeastern U.S. \cite{Doe2004} and 2012 blackout in India \cite{India2012} gave us the evidence about the catastrophic failures which can have acute effects on the modern world. Excluding the natural disasters such as earthquakes, hurricanes and solar flare, the power grid is also vulnerable to terrorists and Electromagnetic (EMP) attacks \cite{Us2010}. Because of this, the utility industry is undergoing through massive grid modernization efforts in effort to tackle the situations like the blackouts discussed above. The massive blackouts came as a wake up call to utilities and consumers alike, climate change, overpopulation and scarcity of natural resources. In reaction, the industry began to strive for a more intelligent power grid called Smartgrid. 
One of many "smart" features of the smart grid is automated microgrids that connect seamlessly with the main grid, and may operate independently when needed. This ensures efficiency, security, reliability, quality and sustainability for energy consumers and producers alike.

A microgrid is defined as a discrete energy system consisting of distributed energy sources (e.g. renewables, conventional, storage) and loads capable of operating independently from the main grid. The important goal of microgrid is to increase the self-sufficiency of energy and independence of system operation in local energy communities.
A microgrid includes generation, a distribution system, consumption and storage, and manages them with advanced monitoring, control and automation systems. A fully-developed microgrid has the capability of automatically disconnecting and operating independently from the main grid. For example, in case of a natural disaster, if the energy service is disrupted from the main grid, the automated controls will cut off the microgrid from the main grid in order to avoid cascading failure. 
Given there is a blackout or cascading failure in the power network, the microgrids separate themselves from the main grid. Or the blackout has itself caused separation of the microgrid from the main power grid. The internal power generating resources balance the demands after the separation. 

The existing class of cyberattacks and EMP attacks are targeted towards the main grid without considering the reliability of the microgrids. Microgrids are safe from the cascading failure effect caused by the above attacks in the main grid. They separate themselves from the main grid and operate independently once they sense disruption or the main grid is cut off. Also a single point attack is not possible given the distributed generation units of the microgrid is local to itself than a central generation as in the case of the main grid. In order to attack the microgrids, only the severe blackouts in the other areas of the smart grid would not have any impact. In the other way around, cyber attacker would not be successful in attacking nodes in  microgrid unless the microgrid is islanded, because they would still keep receiving power from the main grid. In addition, power generation done at a smaller scale is costlier, hence the microgrid's self generation is much costlier than the power provided by the main grid. Taking both the scenarios into the consideration, we demonstrate that the microgrid is still vulnerable to cyber attacks. 

In our approach, we force the microgrids run solely on their own power by islanding them from the main power grid. We achieve this through price modification attack on the entire grid, but with focus on islanding the microgrid i.e. attacking nodes which will lead to failure of edges separating microgrid from the main power grid. The next step involves launching the price modification attack inside each of the microgrids, to fail as many lines as possible and cause cascading failures, leading to massive failure in the smart grid as well as the microgrids connected to the smart grid.  Our contribution to the paper are 1) We have proposed a new cyberattack through price modification against the microgrids. We have structured the attacks given the self sustaining nature of the microgrids. It is a 2 step attack, where first we separate the microgrids from the main grid and in the second step, we attack the load nodes of the microgrid itself.  2) We have ran the experiments on IEEE 14 Bus data along with IEEE 9 Bus data serving as the microgrids.

The rest of the paper is as structured as follows. Section \ref{sec:Powerflow} describes the DC power flow model used in the Smart grid including the microgrid. Section \ref{sec:model} describes the attack model for the Price Modification Attack (PMA). In section \ref{sec:exp}, we evaluate our proposed algorithm and discuss mitigation measures in section \ref{sec:measure}. Section \ref{sec:conc} concludes our paper.

\section{Power Flow} \label{sec:Powerflow}

Power flow in smart grid as well as the microgrid are governed by power laws. We have adopted the widely used model of linearized DC model over the AC model. DC power flow approximation is chosen because it is solved far more quickly and is proved to be accurate under good operating conditions. In order to reduce the running time and make the linearization feasible, the following assumptions are taken into consideration:
\begin{itemize}
	\item Phase angle differences are small and hence we can use $\sin \theta_{ij} = \theta_{ij}$ and $\cos \theta_{ij} = 1$, where $\theta_{ij}$ is the phase angle between the voltages at the two nodes $i$ and $j$.
	\item Lossless lines; i.e. the resistance of the arcs/lines is negligible. 
	\item Voltage profile is kept flat.
\end{itemize}

We briefly describe the linearized or DC power flow model. In the linearized approximation, we are given a power grid represented by a directed graph $G$, where:
\begin{itemize}
	\item Each node $i \in V$ corresponds to either a power generator (i.e., a supply node), or to a load (i.e., a demand node), or to a node that neither generates nor consumes power. The set of generator nodes are denoted by $P$.
	
	\item If node $i$ is a generator, then there are values $0 \le P^{min}_i \le P^{max}_i$. If the generator is operated, then its output must be in the range [$P^{min}_i$,  $P^{max}_i$]; if the generator is not operated, then its output is zero. In general, $P^{min}_i > 0$.
	
	\item If node $i$ is a demand, then the "nominal" demand is given by $D^{nom}_i$. The set of demands or demand nodes is denoted by $D$.
	
	\item The edges $E$ represent power/transmission lines. For each line $(i, j)$, two parameters are given i.e. $x_{ij} > 0$ (the resistance or reactance) and $u_{ij}$ (the capacity).
\end{itemize}

Now, given a set $P$ of operating generators, the linearized power flow is a solution to the system of constraints given in the following set of the equations. For each edge $(i,j)$, $f_{ij}$ represents the power flow on the edge (transmission line) $(i,j)$. In the case where $f_{ij} < 0$, power is effectively flowing from $j$ to $i$. Additionally, the phase angle at node $i$ is given by the variable $\theta_i$. Given a node $i$, $\delta^+(i) (\delta^-(i))$ is the set of lines oriented out of (into) node $i$. 
The power flow equations are given below:
\begin{align}\scriptsize
	\sum_{(i,j) \in \delta^{+}_{i}}f_{ij} - \sum_{(j,i) \in \delta^{-}_{i}}f_{ji} & = 
	\left\{ 
	\begin{array}{l l}
		P_i & \quad i \in P\\
		-D_i & \quad i \in D\\
		0 & \quad \text{otherwise}
	\end{array} 
	\right. \label{pf1}
\end{align} 
\begin{align}\scriptsize
	\theta_i - \theta_j -x_{ij}f_{ij}=0, &&\quad\forall (i,j)\in E\label{pf2}\\
	P_i^{min}  \leq  P_i \leq  P_i^{max}, &&\quad \forall i \in P \label{pf3}\\
	0 \leq  D_j \leq D_j^{nom}, &&\quad \forall j \in D \label{pf4} 
\end{align}

\section{Attack Model and Proposed Algorithms} \label{sec:model}

\begin{figure}[h]
	\centering
	\includegraphics[width=0.5\textwidth]{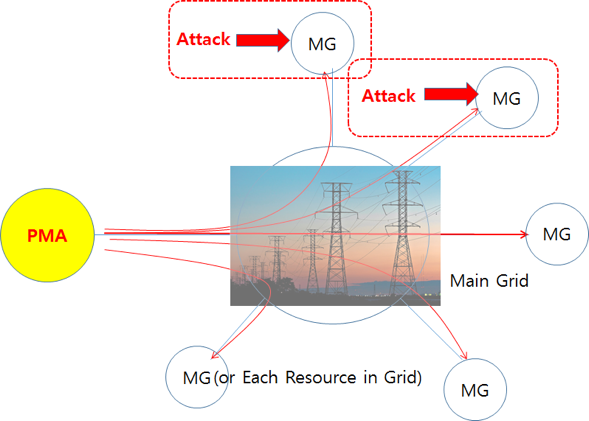}
	\caption{Overview of the PMA.}
	\label{fig:overview}
\end{figure}

In this section, we introduce the two step price modification attack (PMA) in the smart grid to disrupt the services of the microgrid as shown in Fig \ref{fig:overview}. Given the robustness of smart grid, the price modification attack is launched in 2 steps. Step 1 being the process to separate the microgrid from the main grid called as islanding, so that they start operating solely on the local distributed generation sources (which includes PVs, generators and batteries). In step 2, we attack the individual component in the microgrid to make the system go unstable and cause failure of other nodes inside the microgrid.

\subsection{Cascading Failure}

Before moving on to the attack, we discuss the cascading failure model  adopted in the paper. We have adopted the cascading failure described in \cite{Mishra2015, Bernstein} which is an extension of the model as proposed in \cite{Chen}. The total power generation matches the total demand, given the steady state of the power grid. When some lines/arcs are disconnected, or they fail, they are removed from $G$. Following which, the total demand and supply are balanced or adjusted within each island or component by decreasing the load and supply at the various buses. Overall system state is again calculated given the linearized DC power model described in the previous section. The new flows may exceed the capacity and as a result, the corresponding lines will become overheated.  The  outages are modeled by moving average of the power flow $\makeatletter \tilde{f}^t_{ij}$:  $\tilde{f}^t_{ij} = \alpha f_{ij} + (1-\alpha)\tilde{f}^{t-1}_{ij} \makeatother$. Here parameter $\alpha$ used to encode memory so as to model thermal effects. For memoryless system, $\alpha = 1$. For the paper, we use the following rule: line fails if $f_{ij} > u_{ij}$, where $u_{ij}$ is the capacity or flow limit for the transmission line $ij$. The moving average approximates thermal effects, including heating and  cooling from prior states to first order \cite{Anghel}. The algorithm for the cascading failure is given by Algorithm \ref{alg:template}.

\begin{algorithm}[h]
	\scriptsize
	\caption{Cascade Failure Template}
	\label{alg:template}
	\KwData{Connected Power Grid Network $G(V,E)$}
	\KwResult{$S_1$: Lines which failed \\ $S_2$: Nodes which failed}
	\While{Network is not stable}{
		Adjust the total demand to the total supply within each island.\\
		Use equations (1)-(4) to calculate power flows in G. \\
		For all lines computer the moving average 
		$\tilde{f}^t_{ij} = \alpha f_{ij} + (1-\alpha)\tilde{f}^t_{ij} \makeatother$.\\
		Remove all lines that have moving average flows greater than the capacity ($\tilde{f}^t_{ij} > u_{ij}$)  and add to $S_1$.\\
		Add the failed nodes to $S_2$.\\
		If no more line fails, then network is stable, break the loop.
	}
	\textbf{Return} $S_1,S_2$
\end{algorithm} 

\subsection{Islanding the Microgrid}

Given the lower protection and lower sophistication involved in attacking the consumption sector \cite{rad2011} (the other sectors are generation and distribution and control), we focus our attention to this sector.. This class of cyber-intrusion alters the electricity prices through fabricated price signals thereby altering load  at certain grid locations through the Internet and by means of automatic and distributed software intruding agents. In order to island the microgrid, we need to disconnect the main power generation sources from the microgrid. Attacker alters the price of electricity through modification of the price signal sent from the energy distributor over the communication network. This causes disruption in the service from the main grid to the microgrids and thus forcing the microgrid to work in islanded mode. Once the microgrid is in islanded mode, its own power generators are the only sources available to it, which becomes the focus of attack in the second step of the attack.\\

\textbf{Minimum cost to break e* [MCB(e*)]}
{\scriptsize
	\begin{empheq}[box=\fbox]{align}
	\min \quad & \sum_{i \in D}c_i(z_{i}) \label{eq:2obj}\\
	s.t. \quad (1 + k_i) B_i & \geq  D_i \cdot (r_i - z_i \cdot \rho_i) && \forall i \in D \label{eq:2const1} \\ 
	P_{ij} & > u_{ij} && ij = e^* \label{eq:2const3}\\
	z_i & \in [0,1] && \forall i \in D  \label{eq:2const6}
	\end{empheq}
}

Every bus/user $i$ receives the electricity rate $r_i$ from the Internet which we assume to be accessible to the attacker to manipulate or alter. $D_i$ is the total demand at the load node $i$. The total bill is given by $(1 + k_i)B_i$ where $B_i$ represents the user’s targeted billing amount and $k_i$ represents the sensitivity of the user towards billing amount. $k_i = 0$ indicates that the user is not willing to pay anymore than the targeted billing amount. For simplification we allow the $k_i$ with a maximum value of $1$. The line $(i,j)$ fails when the power flow through the transmission line goes over its capacity. A certain portion of the rate is allowed to change for each user given the baseline constraints set up or hard coded by electricity distribution companies. Hence, we have the maximum rate change (MRC) $\rho_i$, which represents the maximum change in the rate that the attacker is allowed. The attacker can choose what percentage of the MRC it wants to change denoted by $z_i$ which lies between 0 and 1. 

Given the automated demand side management, one of the "smart" features of the smart grid, the rate change causes automated increase in the use of the demand for the same household or the company, such as starting up the laundry, more frequent use of the heating/cooling devices, etc. There is a cost associated with the change done by the attacker. Here we consider a linear cost function $c(.)$ for simplicity. The integer programming for calculating the minimum cost for breaking the edge $e^*$ is given by $MCB(e^*)$. After knowing the MCB for each edge, we find the lines that need to broken to separate the microgrids from the main grid, called Islanding the Microgrids (IM) given by Algorithm \ref{alg:im}.

\begin{algorithm}[h]
	\scriptsize
	\caption{Islanding the microgrid (IM)}
	\label{alg:im}
	\KwData{Connected Power Grid Network $G(V,E)$, Budget $A$}
	\KwResult{$S_1$: Lines which failed, $M$ Microgrids that were islanded}
	Initialize current cost $T = 0$\\
	\While{$T<A$}{
		Sort the lines ($E^*$) according to their islanding potential in increasing order\\
		\ForEach{$e^* \in E^*$}{
			\eIf{$T+MCB(e^*) < A$}{
				Break the edge $e^*$\\
				$T = T + MCB(e^*)$\\
				$M = M \cup$ microgrids disconnected from main grid\\
				$E^*$ = $E^*$ - disconnected edges \\
				$S_1$ = $S_1 \cup $ disconnected edges 
			}
			{
				break
			}
		}
	}
	\textbf{Return} $S_1,M$
\end{algorithm} 

\begin{figure*}[ht!]
	\centering
	\includegraphics[width=\textwidth]{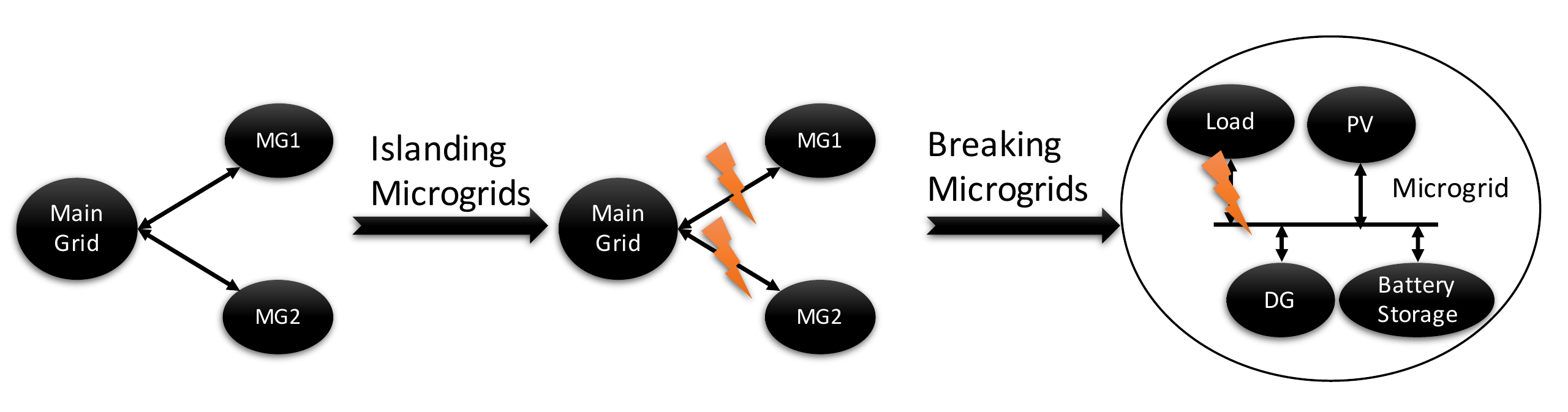}
	\caption{Price Modification Attack showing two steps of attack. Step 1: Islanding Microgrid(IM): Separating the microgrid from the main grid and forcing it to work independently. Step 2: Breaking Microgrid (BM) Failing the load nodes inside the microgrid. In the figure, MG1 and MG2 represent the microgrids 1 and 2 that are connected to the main grid. PV and DG denote Photovoltaic panel and diesel generator.}
	\label{fig:pma}
\end{figure*}

In IM, we first calculate the minimum cost to break (MCB) for all the transmission lines. Islanding potential of a line is given by the number of the microgrids that are separated from the main grid due to the failure of the corresponding line's failure divided by the MCB. This is followed by sorting of all the edges in the increasing order of their potential. We fail the edges in the computed order, given that the constraint of budget $A$ is met. We keep updating the set of disconnected or islanded microgrids $M$ and the set of disconnected edges $S_1$. The temporary budget usage $T$ is also increased by the MCB every iteration where there is a line failure. This process is repeated till we exhaust out of resources or all the microgrids are islanded. 

\subsection{Breaking the Microgrid} 

After we have islanded the microgrids, they start to operate independently from the main grid and its own power generation sources (such as photovoltaic panels, battery storage and diesel generators) ramp up to meet the demand of the load nodes that have been separated from the power grid. This feature ensures the reliability of the microgrid  and power security for the consumers in the microgrid. Although the microgrids are much smaller in nature, the attacker can modify the rates of particular generation units inside the microgrid, leading to the failure of the particular generation unit and leading to the further failures of other generating units in the microgrid ending with complete failure of the microgrid.

\textbf{Breaking the transmission Lines (BL)}
{\scriptsize
	\begin{empheq}[box=\fbox]{align}
	\max \quad & \sum_{e(ij) \in E}y_{ij} \label{eq:1obj}\\
	s.t. \quad (1 + k_i) B_i & \geq  D_i \cdot (r_i - z_i \cdot \rho_i) && \forall i \in D \label{eq:1const1} \\ 
	\sum_{i \in D}c_i(z_{i}) &\leq A &&  \label{eq:1const2}\\
	y_{ij} & < 1 + \frac{f_{ij} - u_{ij}}{u_{ij}} && \forall e(ij) \in E  \label{eq:1const3}\\
	y_{ij} & \in \{0,1\} && \forall e(ij) \in E  \label{eq:1const5} \\
	z_i & \in [0,1] && \forall i \in D  \label{eq:1const6}
	\end{empheq}
}

In this second step of the PMA, that is breaking the microgrid, our aim is to fail as many as loads inside the microgrid. Failure of a load or demand node occurs when it is separated from all the generating resources similar to the islanding process of the microgrid itself. Here we utilize the price modification cyberattack to increase the load of the demand nodes by changing the price information for the generation units. This leads to unbalance and recalculation of power flow in the network and can cause overflow in the circuits and leads to transmission line failures resulting in the break down of the demand nodes. 

However in the microgrid due to the presence of multiple source of energy generation, instead of reducing the Internet price for the load nodes and randomly attacking any power generating device, intuitively the lowest capacity generation device is selected. The rate for the particular generation device is then lowered. Once most of the demand node start using the selected source for power usage, the overall load in the network is manipulated to cause maximum number of failures through Breaking the transmission Lines (BL) algorithm. This continues till the resource gets exhausted. For simplicity we assume a linear cost function $c_i^{gu}$ for modifying the price of the generation unit $i$. The above proposed approach is presented in Algorithm \ref{alg:bm} namely Breaking the Microgrid (BM).

\begin{algorithm}[h]
	\scriptsize
	\caption{Breaking the microgrid(BM)}
	\label{alg:bm}
	\KwData{Connected Power Grid Network $G(V,E)$, Budget $A$}
	\KwResult{S: set of failed nodes inside microgrid}
	Initialize current cost $T = 0$\\
	\While{$T<A$}{
		Sort the generation units ($GU*$) in increasing order of their capacity\\
		\ForEach{$gu_i^* \in GU^*$}{
			\eIf{$T+c^{gu}_{i}(gu_i^*) < A$}{
				Reduce price of the $gu_i$ for all nodes \\
				$BL(gu_i)$ //Breaking the lines with price changes \\
				$GU* = GU* - gu_i$\\
				$S = S \cup$ failed nodes
			}{ break
		}
	}
}
\textbf{Return} $S$
\end{algorithm}

In BM, the temporary budget usage $T$ is check against maximum resource $A$ to verify if resources are available to cause internal demand node failures inside the microgrid. If so, a sorted list of the generation units according to their capacity in increasing order is used while picking the generation source for the corresponding price modification. If the cost $c_i^{gu}$ follows the resource constraint, we alter the price and perform the breaking the transmission lines (BL). This process is repeated till the resources are exhausted and the set of failed nodes $S$ is returned.

\subsection{Price Modification Attack}

We combine the above two steps Islanding the Microgrid and Breaking the Microgrid to launch the Price Modification Attack (PMA) given by Algorithm \ref{alg:pma}. We initiate the PMA with disconnecting the microgrid from the generating source and BM is executed once there is one islanded microgrid. This is continued till we exhaust the maximum resource allocated $A$. 

\begin{algorithm}[h]
	\scriptsize
	\caption{Price modification attack (PMA)}
	\label{alg:pma}
	\KwData{Connected Power Grid Network $G(V,E)$, \\Budget $A$}
	\KwResult{$S1$ Number of node failures; $S2$ Number of microgrids islanded; $S3$ Number of node failures inside the microgrid}
	\While{max resource $A$ not utilized}{
		Islanding the Microgrids (IM) \\
		\If{$\exists$ Islanded microgrid}{
			Breaking the Microgrids (BM) \\
		}
	}
	Return $S1,S2,S3$
\end{algorithm}

\section{Experimental Evaluation} \label{sec:exp}
\begin{figure}[h]
	\centering
	\includegraphics[width=0.45\textwidth]{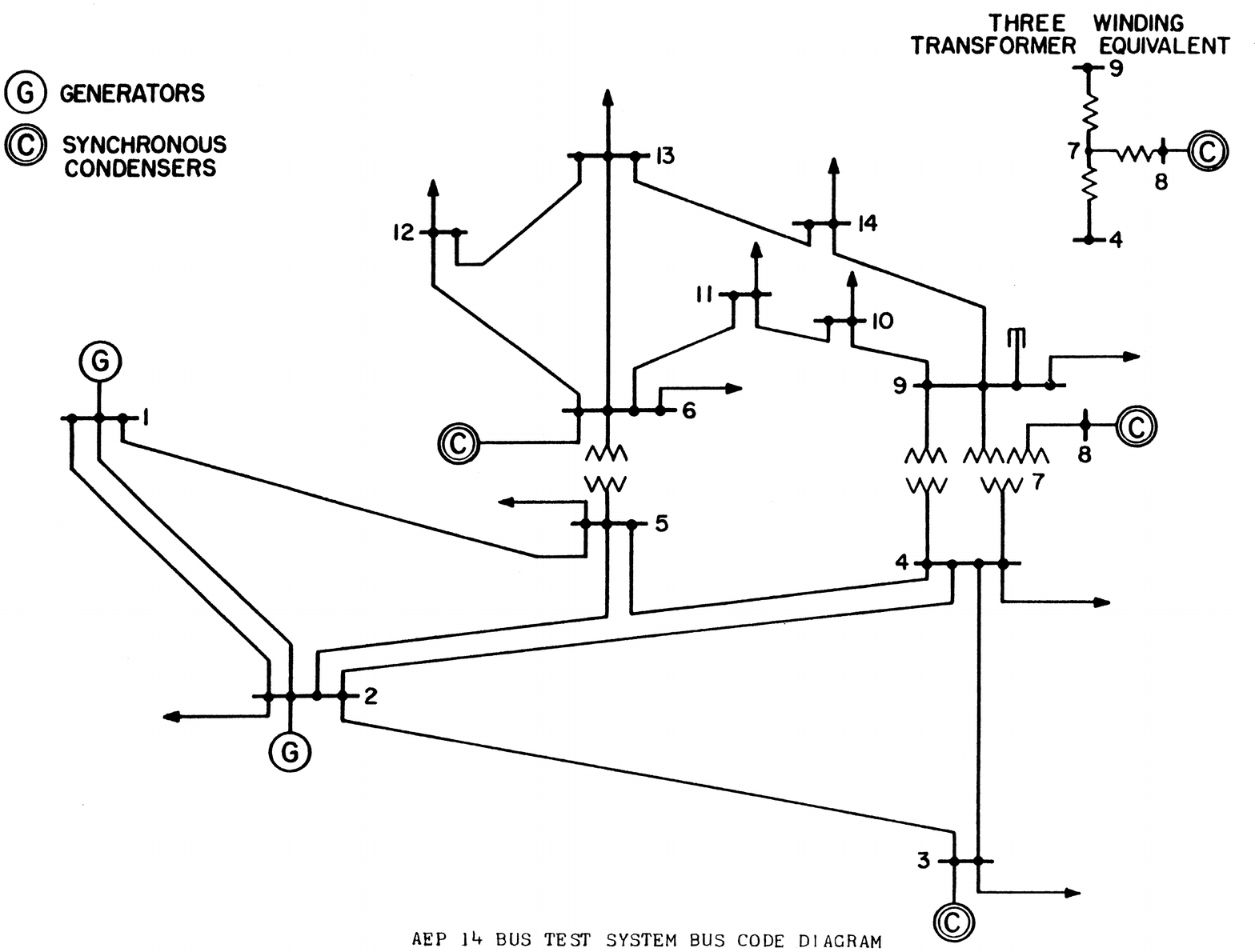}
	\caption{IEEE 14 bus system data. This represents our basic smart grid, we assume the location of the microgrids to be at bus number 13 and 14. \cite{ieeedata}}
	\label{fig:ieee14}
\end{figure}

\begin{figure}[h]
	\centering
	\includegraphics[width=0.4\textwidth]{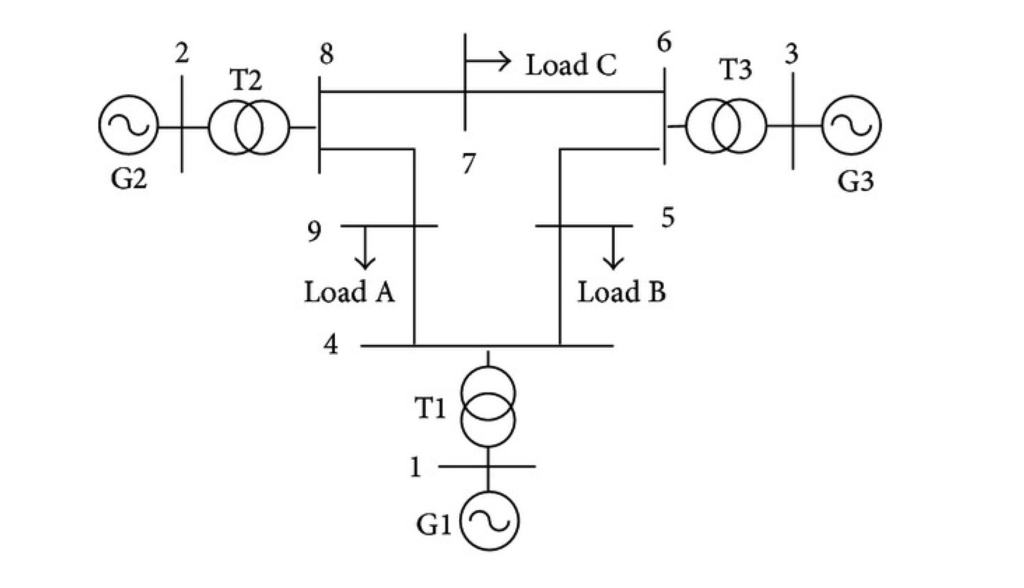}
	\caption{IEEE 9 bus system data. This consists of 3 generators such as G1, G2, and G3; and 3 loads namely Load A, Load B and Load C.}
	\label{fig:ieee9}
\end{figure}

In this section, we evaluate the efficiency of the different algorithms we proposed. In the experiments reported in this section we used a 3.0 GHz Xeon machine with 2 MB L2 cache and 12 GB RAM. All experiments were run using a single core. For the power system simulation, we use the matlab package called MatPower \cite{matpower}. 50 runs of each cycle was run and averaged for consistency. We use the random algorithm as a baseline to compare our proposed algorithms. In order to represent the microgrid system we take IEEE 14 bus system data as the smart grid as shown in Fig \ref{fig:ieee14} and set up microgrids on bus 13 and 14. To represent the microgrid architecture, we use the IEEE 9 bus test system with 3 generators and 3 loads as shown in Fig. \ref{fig:ieee9}. By default we keep the system at  the line capacity reduced by $60\%$ and maximum resource at $20\%$ and microgrid load is kept at the default. Node or load failure occurs when a particular load gets cut off from the all power generating sources. Microgrid is islanded when it is separated from the power generating source in the main grid and thereafter starts operating independently. 

In all the experiments the same approach was employed: first, we calculated the optimal powerflow in the main grid with the preset parameters \cite{ieeedata}. In case of load imbalance due to cascading failure or price modification; both the demand and total power were made equal and power flow was recalculated. The line was failed if there was any overflow i.e. $P_{ij} > u_{ij}$. 

\begin{center}
	
	\begin{figure*}[ht]
		\centering
		\begin{subfigure}[b]{\textwidth}
			\includegraphics[width=0.32\textwidth]{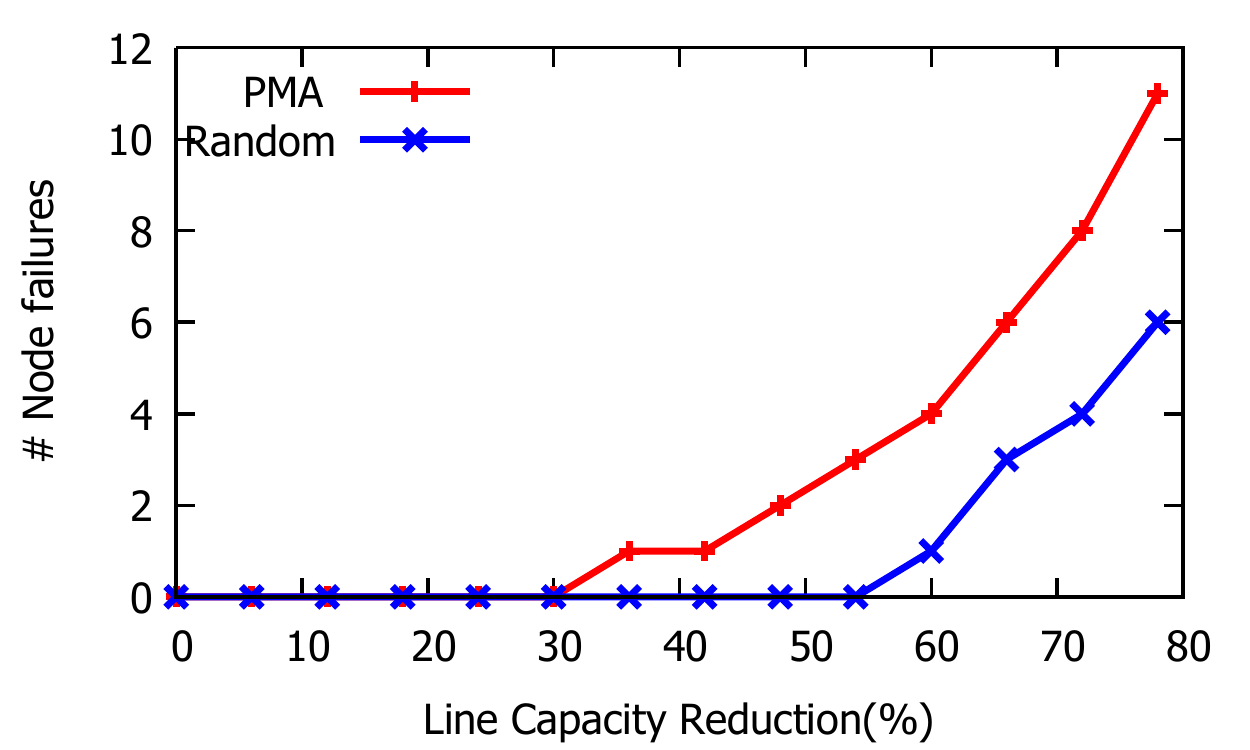}
			\includegraphics[width=0.32\textwidth]{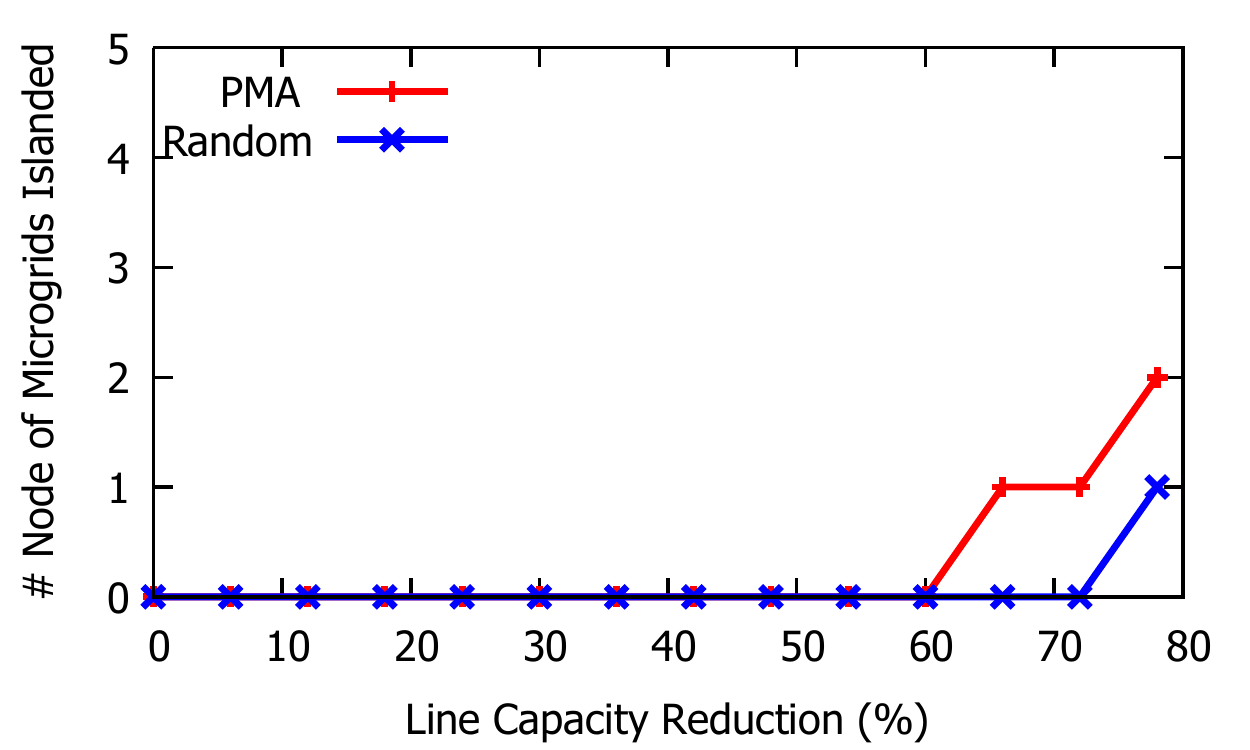}
			\includegraphics[width=0.32\textwidth]{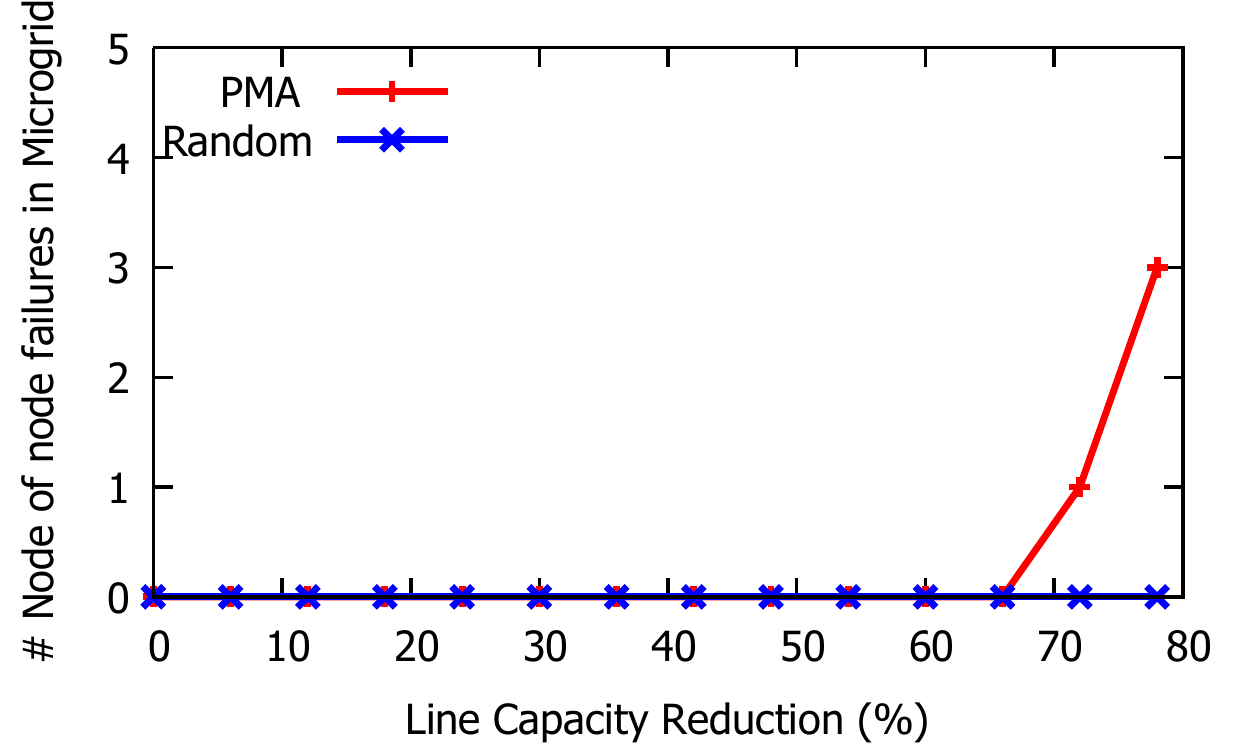}
			\caption{Line Capacity}
			\label{fig:lines_c}
		\end{subfigure}
		\begin{subfigure}[b]{\textwidth}
			\includegraphics[width=0.32\textwidth]{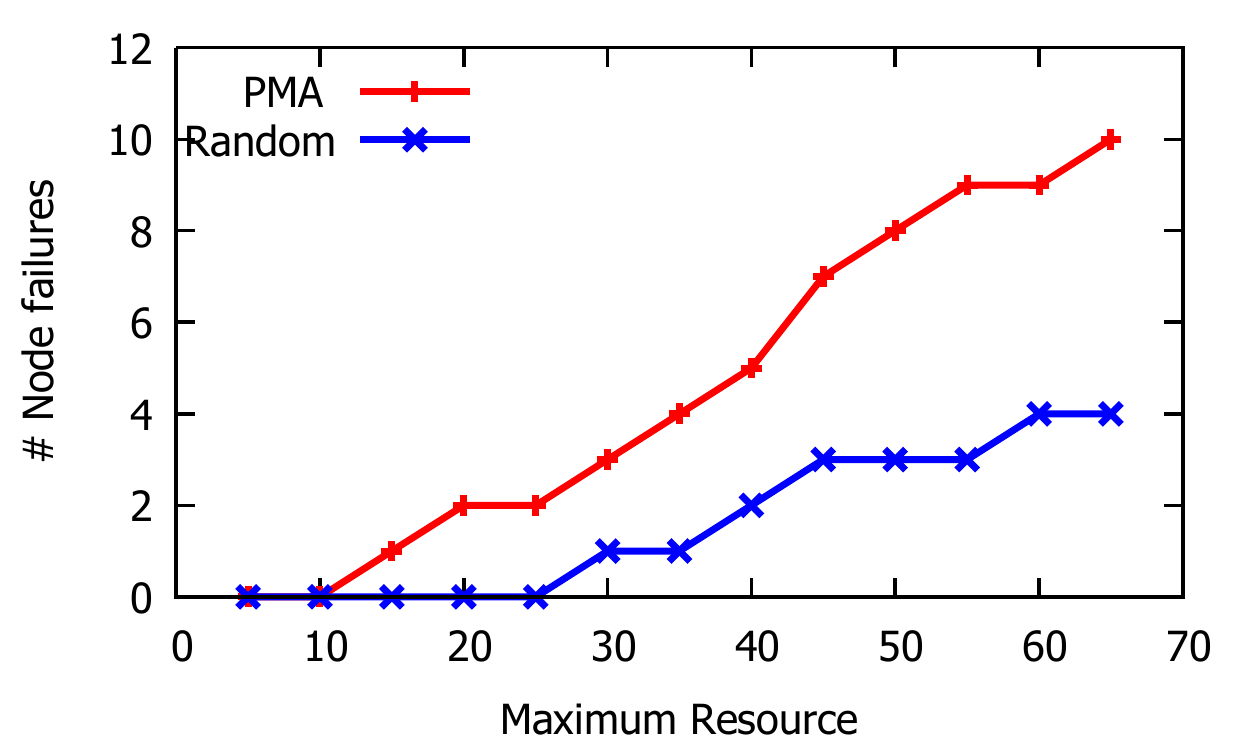}
			\includegraphics[width=0.32\textwidth]{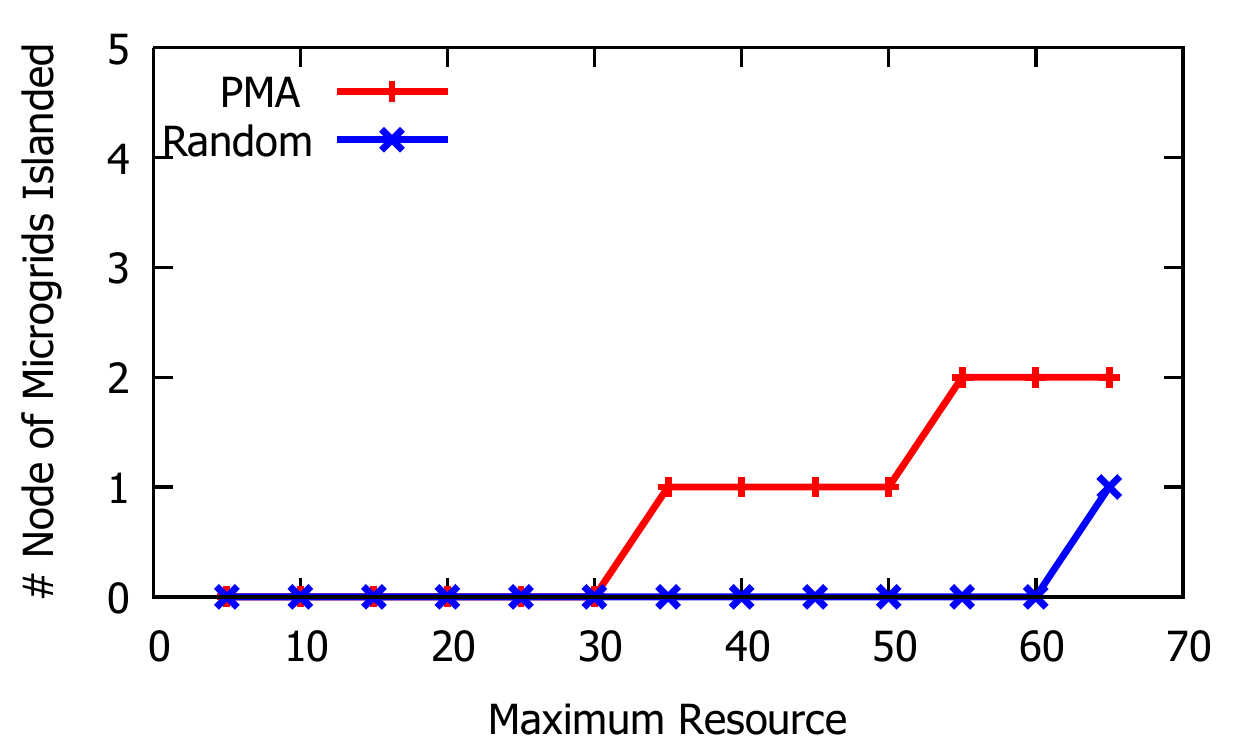}
			\includegraphics[width=0.32\textwidth]{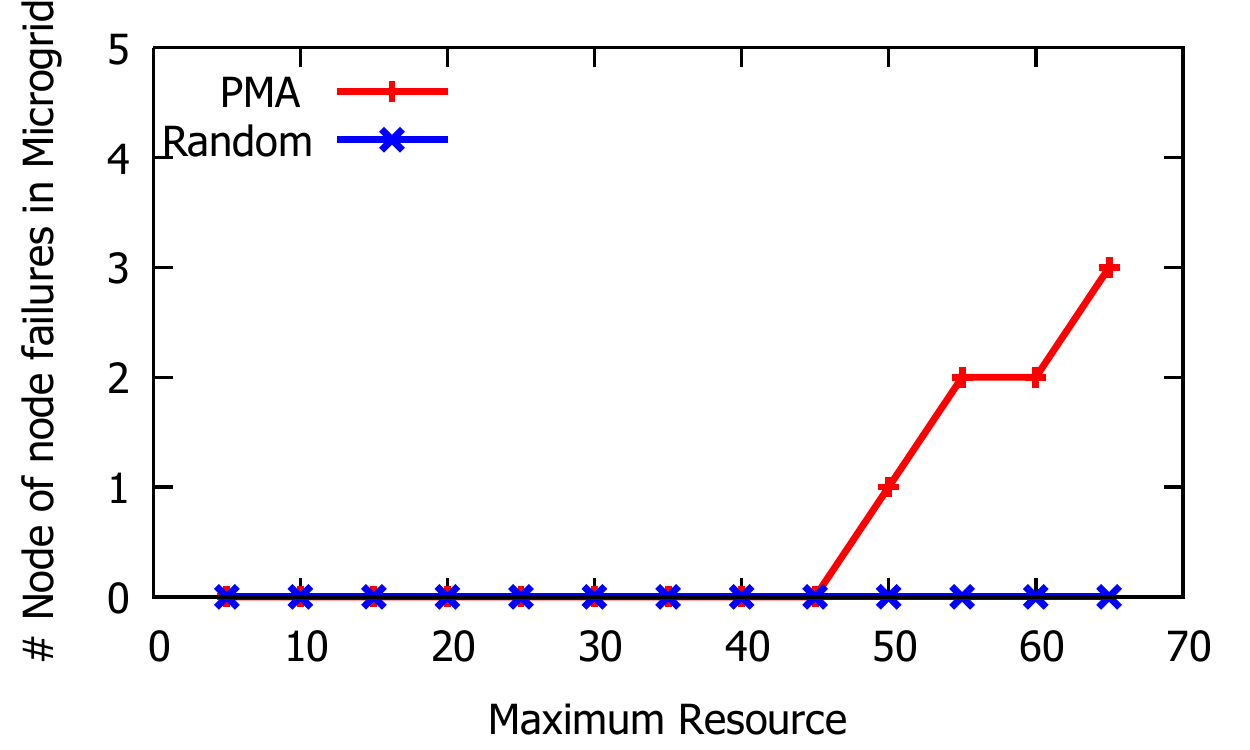}
			\caption{Maximum resource (\%)}
			\label{fig:lines_r}
		\end{subfigure}
		\begin{subfigure}[b]{\textwidth}
			\includegraphics[width=0.32\textwidth]{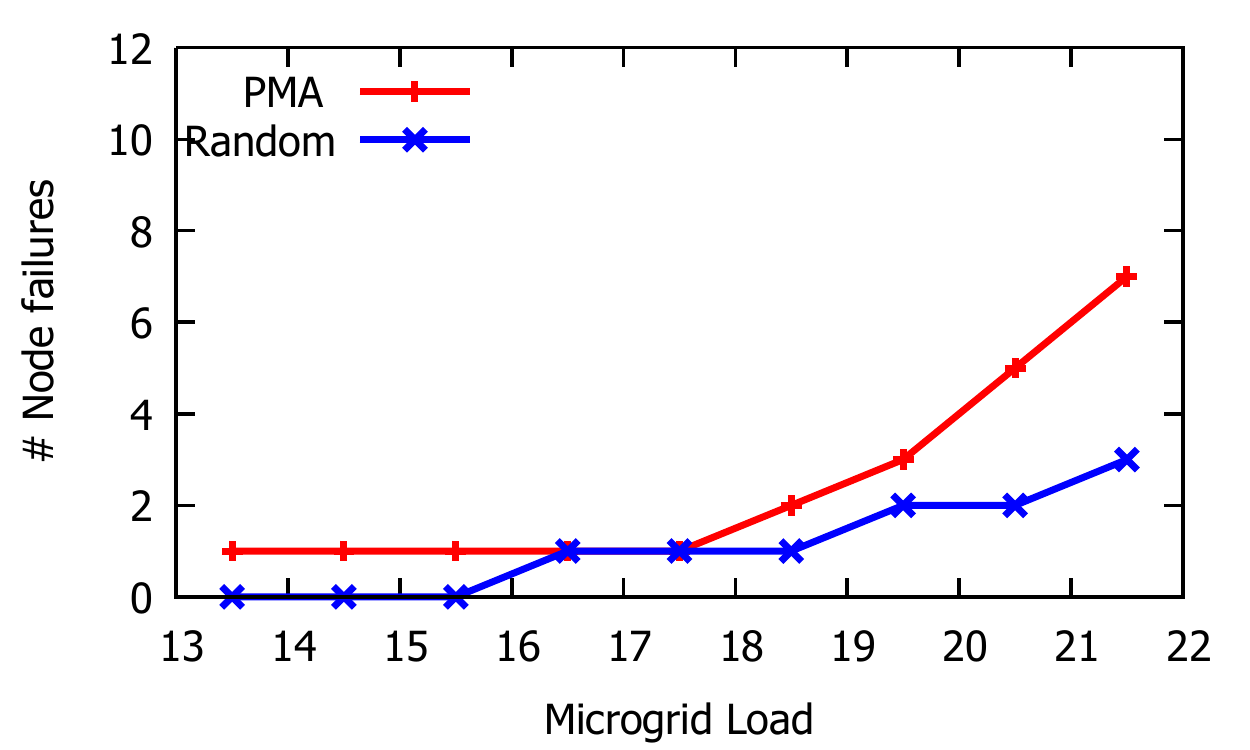}
			\includegraphics[width=0.32\textwidth]{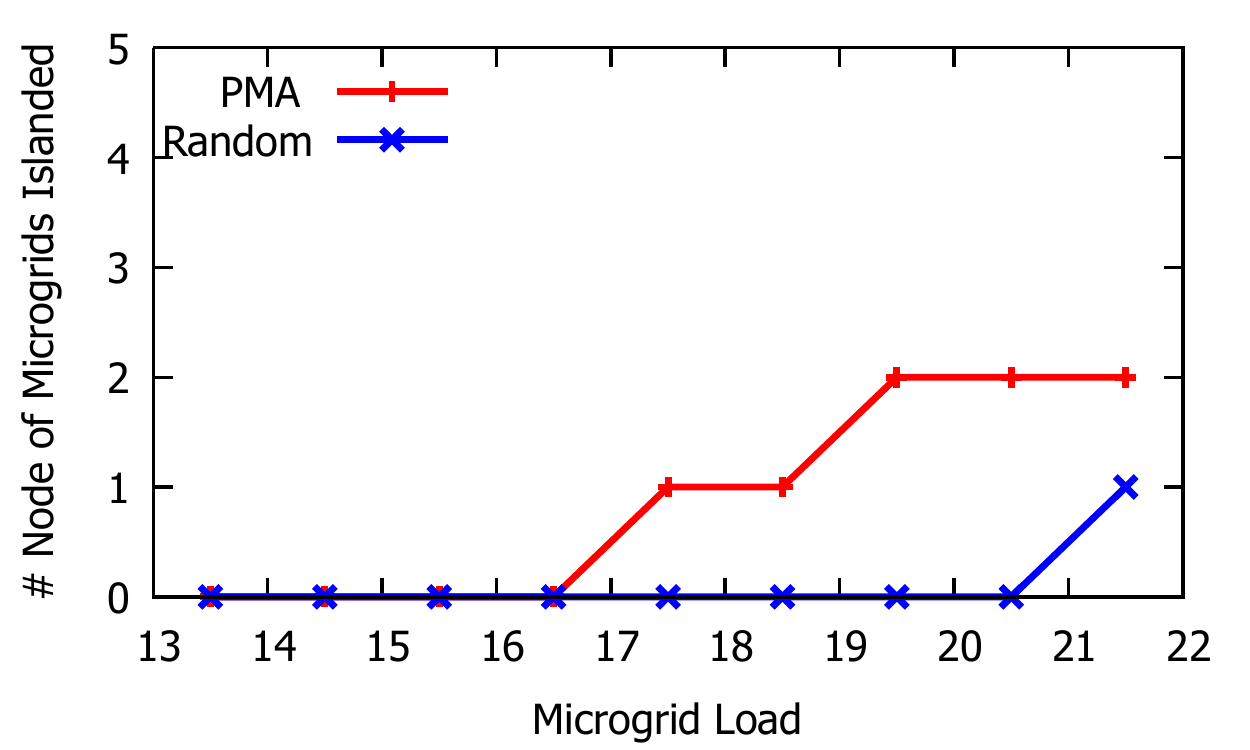}
			\includegraphics[width=0.32\textwidth]{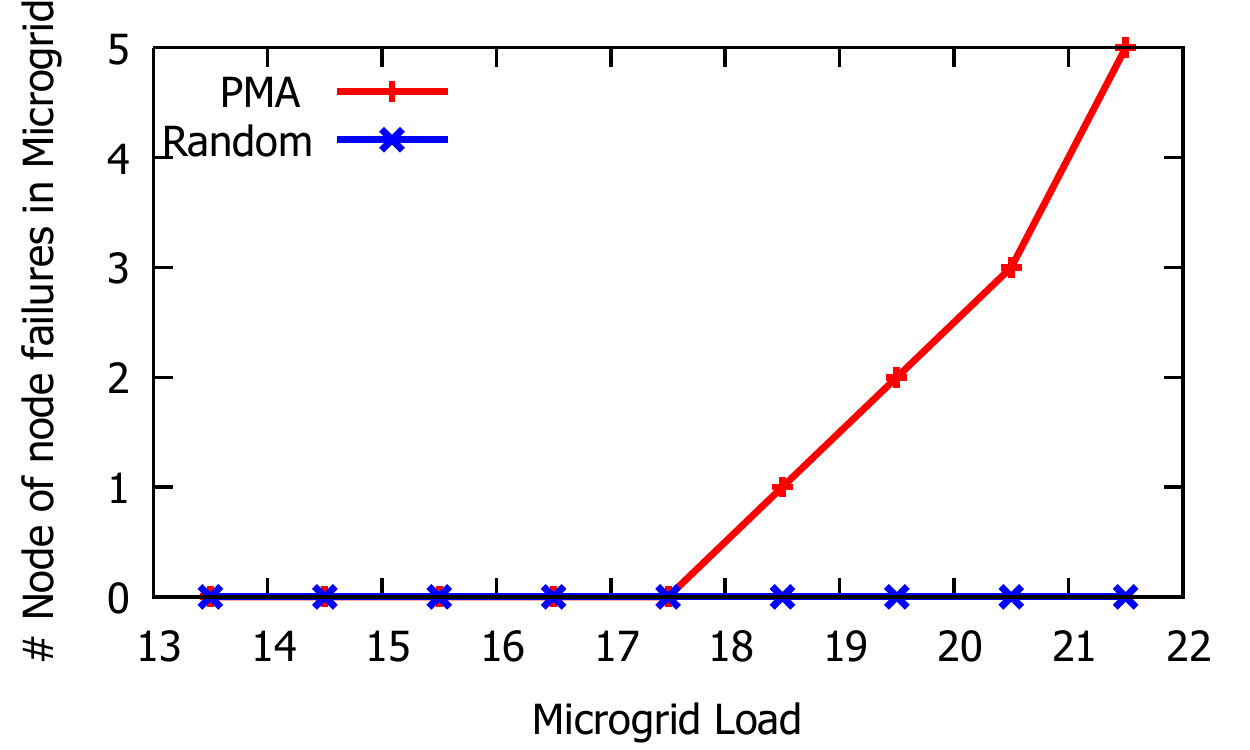}
			\caption{Microgrid Load}
			\label{fig:lines_a}
		\end{subfigure}
		\caption{Evaluation of Price Modification Attack with respect to various grid parameters such as line capacity, maximum resource and microgrid load.}\label{fig:lines}
		
	\end{figure*}
\end{center}

In the Fig. \ref{fig:lines_c}, we evaluate the PMA by varying the line capacity in the grid. By reducing the line capacity, the power flow in the transmission gets closer to capacity of the network, thus increasing the stress level of the grid. As the capacity gets reduced, it becomes much easier to attack and increase the power overflow the transmission line's capacity. Here we reduce the line capacity by percentage from $0$ to $78$ and evaluate PMA algorithm on the grid with respect to the variance. As we can verify from the experiment, PMA algorithm performs really well not just in the overall node failures, but is also to separate the microgrids from the main grid and failing the nodes inside the microgrid. Although the random algorithm is blind and not able to island the microgrids as efficiently and performs really worse in failing the loads inside the microgrid.

Then we evaluate the PMA with respect to change in the maximum resource available to the attacker as shown in Fig \ref{fig:lines_r}. We vary maximum resource $A$ from $0\%$ to $65\%$ and verify the effect of PMA and random algorithm in the grid. Attacker gains more means to attack the grid and more loads to reduce the price information such that the overall load in the grid increases with the increase in the maximum resource. Hence, the rise in the number of load failures, islanded microgrids and load failures inside the microgrids. 

Here we test the reliability of the microgrids against PMA given the variance in overall variance of microgrid loads. We vary the load from $13.5$ units (default) to $21.5$ units. As the total load of the microgrids increase, it becomes more vulnerable to PMA, given the increased stress the transmission lines from the generating source to the microgrids. This works in the favor of the PMA and after the islanding, the PMA proceeds with breaking the microgrid and fails much more nodes than the random algorithm.  

\section{Mitigation Measures}\label{sec:measure}

The protection mechanisms from the Internet based price modification attack could be accomplished by protection of price and command signals; smart meter and data center protection; load shedding and load relocation ;and attack detection and learning demand patterns \cite{rad2011}.
Based on the type of load and defense mechanism, full load protection, i.e., protecting all vulnerable loads, can be a significant expense. Assuming that protecting every node in the network against price modification attack is neither feasible nor economical, we suggest the approach of the greedily choosing the most critical nodes in the network. These critical nodes are found by the algorithms IM and BM whose electricity prices are modified to bring maximum failure in the network. Protection of critical nodes found in IM algorithm restricts attacker to manually separate the microgrid from the main grid. And protection of critical nodes found in BM algorithm restricts attacker to fail edges inside the microgrid when it starts operating independently. 

\section{Conclusion} \label{sec:conc}
This paper deals with price modification cyberattack on the microgrids as a part of the smart grid. The proposed method is applied to the test system, in order to demonstrate its applicability. The proposed 2-step PMA on the microgrid involves disconnecting the microgrid from the main grid in the first step in order to force the microgrid to operate independently, and in the second step, we try to the fail the nodes inside the microgrid itself. The effect of PMA under various parameters of the power grid network such as line capacity, maximum resource and microgrid load are presented and discussed.
In the future work we would like to consider the multiple variations of the microgrid i.e. AC and DC microgrids, eventually which follow different. We would also consider the impact of both active and reactive power flow and their role in stability of the network or rather instability of the network. 

%
%
%




\end{document}